\documentclass[aps,pra,twocolumn,nofootinbib,10pt]{revtex4-1}
\usepackage{amsmath,amsfonts,amssymb}
\usepackage{siunitx}
\usepackage{graphicx}

\usepackage{tikz}
\usetikzlibrary{backgrounds,fit,decorations.pathreplacing}

\newcommand{\beq}{\begin{equation}}
\newcommand{\eeq}{\end{equation}}
\newcommand{\Z}{\ensuremath{\mathbb{Z}}}
\newcommand{\diag}{\ensuremath{\text{diag}}}

\newcommand{\ket}[1]{\ensuremath{|#1\rangle}}
\newcommand{\bra}[1]{\ensuremath{\langle#1|}}
\newcommand{\ketbra}[2]{\ensuremath{\ket{#1}\!\bra{#2}}}
\newcommand{\dg}{\ensuremath{\dagger}}
\newcommand{\Hc}{\ensuremath{\textrm{H.c.}}}
											
\newcommand{\vac}{\ensuremath{\ket{\text{vac}}}}
\renewcommand{\H}{\ensuremath{\mathsf{H}}}
\newcommand{\CC}{\ensuremath{\mathsf{C}}}
\newcommand{\cp}{\ensuremath{\textsc{cp}}}
\newcommand{\cphase}{\textsc{cphase}}
\newcommand{\enc}{\textsc{enc}}
\newcommand{\swap}{\textsc{swap}}

\newcommand{\BH}{Bose--Hubbard}

\newcommand{\onsite}{\ensuremath{U}}
\newcommand{\hop}{\ensuremath{J}}
\newcommand{\applied}{\ensuremath{V}}
\newcommand{\unitary}{\ensuremath{\hat{U}}}
\newcommand{\Ham}{\ensuremath{\hat{\mathcal{H}}}}
\newcommand{\Hada}{\ensuremath{\hat{H}}}
\newcommand{\numop}{\ensuremath{\hat{n}}}
\newcommand{\cop}[1]{\ensuremath{\hat{c}_{#1}^{\vphantom{\dg}}}}
\newcommand{\cdop}[1]{\ensuremath{\hat{c}_{#1}^{\dg}}}

\newcommand{\rot}{\ensuremath{\hat{R}}}

\newcommand{\comp}[1]{\ensuremath{\underline{#1}}}
\newcommand{\cket}[1]{\ensuremath{\ket{\comp{#1}}}}

\tikzstyle{qubit} = [circle,shading=ball, ball color=black!80!white,
											minimum size=0.75cm]
\tikzstyle{vertex}=[circle, draw, fill=black,
                 minimum size=3pt, inner sep=0pt]
\tikzstyle{smlvtx}=[circle, draw, fill=black,
                 minimum size=2pt, inner sep=0pt]
\newcommand{\railTimestep}[3]{
  \foreach \y / \lbl in {0/1,1/2,2/3,3/4} {
    \node[smlvtx] ({v#3\lbl}) at (#1,#2+3-\y) {};
  }
  \foreach \a / \lbl in {0/5,1/6,2/7,3/8,4/9,5/10} {
    \node[smlvtx] at ({#1+2*cos(180+\a*36)},{#2-1+2*sin(180+\a*36)})
    		({v#3\lbl}) {};
  }
	\draw ({v#35})
		.. controls ++(-1,0.5) and ++(-1,-0.5)
		.. ({v#35});
	\draw ({v#310})
		.. controls ++(1,0.5) and ++(1,-0.5)
		.. ({v#310});
	\draw ({v#37})
		.. controls ++(-0.5,-1) and ++(0.5,-1)
		.. ({v#37});
	\draw ({v#38})
		.. controls ++(-0.5,-1) and ++(0.5,-1)
		.. ({v#38});
  	\begin{pgfonlayer}{background}
		\node[
			inner sep=3mm,
			draw=gray, fill=gray!15, rounded corners=5,
			fit = (v#31) (v#35) (v#37) (v#310)
		] {};
    \node[
    		inner sep=2mm,
    		draw=gray, fill=gray!8, rounded corners=3,
    		fit = (v#31) (v#34)
		] {};
  \end{pgfonlayer}
}
\newcommand{\latticeTimestep}[3]{
  \foreach \x in {1,2} {
  	  \node at ({#1+2*(\x-1.5)}, {#2-1}) {$\scriptstyle(\x)$};
    \foreach \y / \lbl in {0/1,1/0} {
      \node[vertex] ({site#3\x\lbl})
        at ({#1+2*(\x-1.5)}, {#2+2*\y}) {};
    }
  }
}

\newcommand{\swapGraph}[3]{
	\node[draw=gray, fill=gray!8, inner sep=0.4cm, circle] at (#1,#2) {};
	\node[smlvtx] (v#3-1100) at (#1-1,#2+1) {};
	\node[smlvtx] (v#3-1001) at (#1+1,#2+1) {};
	\node[smlvtx] (v#3-0110) at (#1-1,#2-1) {};
	\node[smlvtx] (v#3-0011) at (#1+1,#2-1) {};
	\node[smlvtx] (v#3-2000) at (#1-1-2*0.707,#2+1+2*0.707) {};
	\node[smlvtx] (v#3-0200) at (#1-1+2*0.707,#2+1+2*0.707) {};
	\node[smlvtx] (v#3-0020) at (#1+1-2*0.707,#2-1-2*0.707) {};
	\node[smlvtx] (v#3-0002) at (#1+1+2*0.707,#2-1-2*0.707) {};
	\node[smlvtx] (v#3-1010) at (#1-3,#2-1) {};
	\node[smlvtx] (v#3-0101) at (#1+3,#2+1) {};
	\tikzstyle{edge}=[draw];
	\draw[edge] (v#3-2000) -- (v#3-1100) -- (v#3-0200);
	\draw[edge] (v#3-1001) -- (v#3-0101) -- (v#3-0110)
	        -- (v#3-1010) -- (v#3-1001);
	\draw[edge] (v#3-0020) -- (v#3-0011) -- (v#3-0002);
	\draw[edge] ({v#3-2000})
	  .. controls ++(1,1.5) and ++(-1,1.5)
	  .. ({v#3-2000});
	\draw[edge] ({v#3-0200})
	  .. controls ++(1,1.5) and ++(-1,1.5)
	  .. ({v#3-0200});
	\draw[edge] ({v#3-0020})
	  .. controls ++(1,-1.5) and ++(-1,-1.5)
	  .. ({v#3-0020});
	\draw[edge] ({v#3-0002})
	  .. controls ++(1,-1.5) and ++(-1,-1.5)
	  .. ({v#3-0002});
	\node (bl#3) at (#1-3, #2-1-2*0.707-1) {};
	\node (tr#3) at (#1+3, #2+1+2*0.707+1) {};
}

\begin{document}

\title{Bose--Hubbard model for universal quantum walk-based
computation}
\author{Michael S.\ Underwood}
\author{David L.\ Feder}
\email[Corresponding author: ]{dfeder@ucalgary.ca}
\affiliation{Institute for Quantum Information Science,
University of Calgary, Alberta T2N 1N4, Canada}

\date{\today}

\begin{abstract}
We present a novel scheme for universal quantum computation based on
spinless interacting bosonic quantum walkers on a piecewise-constant
graph, described by the two-dimensional \BH\ model.
Arbitrary $X$ and $Z$
rotations are constructed, as well as an entangling two-qubit
\cphase\ gate and a \swap\ gate.
Quantum information is encoded in the
positions of the walkers on the graph,
as in previous quantum walk-based proposals for universal quantum
computation, though in contrast to prior schemes
this proposal requires a number of vertices only linear in the
number of encoded qubits.
It allows single-qubit measurements to be performed in
a straightforward manner with localized operators, and can make use
of existing quantum error correcting codes either directly within
the universal gate set provided, or by extending the lattice to a
third dimension. We present an intuitive example of a logical
encoding to implement the seven-qubit Steane code.
Finally, an
implementation in terms of ultracold atoms in optical lattices
is suggested.
\end{abstract}

\maketitle

\section{Introduction}

Quantum walks have proven to be a fruitful alternative
to the quantum circuit model for the construction
and description of quantum information processing tasks
\cite{Kempe2003-CP44-307}.
The framework they provide has allowed for the construction of
efficient quantum algorithms
\cite{Ambainis2003-IJoQI1-507},
including some previously unknown under other models
\cite{Ambainis_FOCS2007,Childs2003,Childs2009-ToC5-119,Cleve2008,Farhi2008-ToC4-169,Reichardt2008}
and alternative formulations of Grover's search
\cite{Santha2008}
as well as reproductions of other known results
\cite{Magniez2005,Ambainis2007-SJC37-210}.
Furthermore, it has since been shown that quantum walks
are universal for quantum computation.
Three distinct models
have been described, based on continuous-time quantum walks
\cite{Childs2009-PRL102-180501}, discrete-time quantum walks
\cite{Lovett2010-PRA81-42330},
and what we refer to as the discontinuous walk,
a hybrid method in which a continuous-time quantum walker takes
discrete steps across a set of time-varying graphs, undergoing
perfect state transfer through a subset of the graph at each stage
\cite{Underwood2010-PRA82-42304}.
In each case, the use of a single walker to encode $n$ qubits
leads to a graph with $\Omega(2^n)$ vertices.

One straightforward adaptation that allows for the exponentially
growing Hilbert space required to encode $n$ qubits
without necessitating a similarly sized graph
is to employ multiple quantum walkers.
As a step in that direction,
a framework for the description of multiple non-interacting
discrete-time quantum walkers, distinguishable or not,
has been put forth \cite{Rohde2011-NJoP13-13001}.
It is straightforward to see though that if there are
multiple walkers on a graph yet they do not interact, then there is
no meaningful difference from the situation of multiple distinct
copies of the single-walker situation. Non-interacting walkers
evolve independently of each other, so while they can be useful if
many runs are required to build up statistics of the output state,
no new dynamics can be present.

In the discrete-time quantum walk, effective interactions between
otherwise non-interacting walkers can be induced through the sharing
or swapping of coins \cite{Xue2012-PRA85-22307}, and entanglement
can arise between a single walker and its coin
\cite{Carneiro2005-NJP7-156}.
Continuous-time quantum walks on the other hand have no coin degree
of freedom, so must rely on inter-walker interactions to introduce
additional dynamics and generate entanglement.
There is evidence that on a given graph,
two interacting continuous-time walkers are more computationally
powerful than either
a single walker or two non-interacting ones when applied to the
graph isomorphism problem \cite{Gamble2010-PRA81-52313}.

For the case of the continuous-time quantum walk,
the \BH\ model \cite{Fisher1989-PRB40-546}
provides a natural method with which to describe interacting bosonic
walkers,
and its relationship to a single continuous-time quantum walker
has been discussed in the context of particle transport and
entanglement generation
\cite{Romero-Isart2007-JoPAMaT40-8019}.
The model is well studied in the realm of condensed-matter physics;
in particular it describes strongly correlated systems well
\cite{Bruder2005-AdP14-566},
providing an excellent description of features
such as the transition between the superfluid and Mott-insulator phases,
as well as having been implemented and well controlled experimentally
\cite{Greiner2002-N415-39}.
Furthermore the standard \BH\ Hamiltonian does not address
internal degrees of freedom, allowing us to explicitly consider
the encoding of quantum information within position states, in the
spirit of quantum walks, rather than a more conventional encoding
in spin states.

Standard quantum walk-based schemes for quantum computation
employ a position-based encoding. This results in a spatially
delocalized qubit under circumstances where the quantum walk
formalism can be directly mapped to the circuit model.
In this work, we assign
a pair of vertices to each bosonic walker,
corresponding directly to the computational states $\cket{0}$
and $\cket{1}$.
Gates are implemented by making instantaneous changes
to the graph supporting
the walkers,
and we explicitly require couplings to doubly occupied states
in which two walkers interact on one vertex, despite the fact
that such states do not encode qubits.
The graph is fixed except at the instants
of change, so continuous dynamical control is not required.
The main idea is to design
a discrete sequence of graphs, such that
at the beginning and end, the spatial separation of the walkers
is maintained with one walker per pair of vertices, yet
during their evolution the walkers have
the opportunity to interact whenever
two walkers occupy one vertex. This allows states with
doubly occupied vertices to be harnessed as an advantage in
constructing logical gates, rather than having to be strictly avoided
as a source of decoherence during gate operations as in previous
proposals \cite{Mompart2003-PRL90-147901}.

In this work we provide an important correspondence between the
discontinuous evolution of interacting indistinguishable particles
on polynomial-sized graphs and
that of a single discontinuous quantum walker
on an exponentially larger graph.
That is, a number of walkers linear in
the number of qubits to be simulated $n$, walking on a vertex set
of a size also linear in $n$,
has the same power as a quantum walk with a constant number of
walkers (in particular, one walker) on a graph with
a number of vertices exponential in $n$.

The structure of this article is as follows.
In Sec.~\ref{sec:PositionStates} we briefly review the \BH\
Hamiltonian, then define an encoding of
computational basis states within a subset of the position states
available to the bosons in the system it governs, and describe a set of
physical operations
that result in a universal set of one- and two-qubit gates
on the computational space.
We then describe in Sec.~\ref{sec:MeasureQECC}
how single-qubit measurements can be made on encoded qubits,
and provide an example scheme for adding a layer of quantum error
correction.
Finally in Sec.~\ref{sec:Implementation} we discuss the possibility of a
proof-of-concept implementation of our scheme, before providing
some concluding remarks in Sec.~\ref{sec:Conclusions}.

\section{Positions as computational states}\label{sec:PositionStates}

\subsection{The \BH\ Hamiltonian}

We use the \BH\ model 
to describe multiple interacting continuous-time quantum walkers
on a graph as spinless bosons hopping on a
lattice, with on-site interactions.
The graph $G$ is defined by a set of vertices $V$
and associated edges $E\subseteq V\times V$. The vertices
correspond to the lattice sites, while the edges indicate
allowed tunnelings.
The \BH\ Hamiltonian is
\beq
	\Ham
	=
	-\hop\sum_{\langle i,j\rangle}
		\cdop{i}\cop{j}
	+ \frac{\onsite}{2}\sum_i \numop_i(\numop_i-1).
\eeq
Here $\cdop{i}$ creates a particle on vertex $i$ and
$n_i\equiv \cdop{i}\cop{i}$
is the number operator on said vertex.  The first sum runs over neighboring
vertices $\langle i,j\rangle$ such that $(i,j)\in E$ is an edge of the
lattice graph.
\hop\ is the tunneling amplitude between
adjacent sites, and \onsite\ is the on-site interaction strength between
the particles.

We refer to the graph $G$, on which multiple quantum walkers
appear, as the \emph{primary} graph.
This primary graph and the $n$ walkers evolving on it comprise a
physical system that can be described by a set of states and allowed
transitions between pairs of them. One can consider each of these
states as corresponding with a vertex in a \emph{secondary} graph,
in which each allowed transition appears as an edge.
The evolution of $n$ walkers
on the primary graph then maps exactly onto the evolution of
a single walker on the secondary graph.
Note that the number of vertices in the secondary graph
is at least exponentially larger than the number of
vertices in $G$.

\begin{figure}
	\begin{tikzpicture}[scale=1]
    \foreach \j in {0.5,1.5} {
      \foreach \i in {0.5,1.5,2.5} {
        \node[style=vertex] at (\i,\j) {};
      }
    }
	  \node[style=qubit, opacity=0.9] at (0.5,1.5) {};
	  \node[style=qubit, opacity=0.9] at (1.5,0.5) {};
	  \draw[style=qubit, opacity=0.5, white] (2.5,1) ellipse (0.25cm and 0.75cm);
	  \node at (-1,1.5) {$x=0,\ \ket{0}\rightarrow$};
	  \node at (-1,0.5) {$x=1,\ \ket{1}\rightarrow$};
	  \node at (0.5,-0.125) {$i=1$};
	  \node at (1.5,-0.125) {$i=2$};
	  \node at (2.5,-0.125) {$i=3$};
	\end{tikzpicture}
	\caption{\label{fig:latticeExample}%
		Cartoon representation of six vertices confining
		$n=3$ particles whose positions
		encode a computational state of the form
		$\ket{\psi}=\ket{0}\otimes\ket{1}\otimes\ket{+}$.
		The physical Hilbert space \H\ for this setup is
		56-dimensional.
		The Hamiltonians we prescribe
		couple the encoded
		8-dimensional computational
		space \CC\ to a total of 24 of the 56 basis states.
	}
\end{figure}
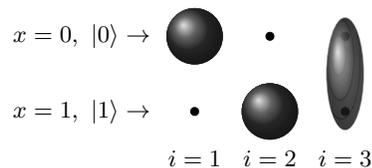

We make use of $n$ bosonic quantum walkers to encode $n$ qubits.
On a primary graph of $2n$ vertices we assign a pair of vertices to each
qubit, and for simplicity visualize the vertices as being arranged
in a $2\times n$ grid structure; see Fig.~\ref{fig:latticeExample}
for a cartoon example with $n=3$.
Label the vertices $v_{i,x}\in V$, with $i\in\{1,\ldots,n\}$ and
$x\in\{0,1\}$, and
let $\ket{0_{i,x}}$ be the vacuum state on vertex $v_{i,x}$ in row $x$
of column $i$. The vacuum state of the system is then
\beq
	\vac
	=
	\ket{0_{1,0}0_{1,1}\cdots0_{n,0}0_{n,1}},
\eeq
which when convenient can also be expressed as
\beq
	\vac
	=
	\ket{00}_1\cdots\ket{00}_n,
\eeq
where the first (second) entry in each pair is understood
to correspond with row 0 (1) of the array.
A walker is created on vertex $v_{i,x}$ by the operator
$\cdop{i,x}$.

The physical Hilbert space \H\ of $n$ bosons confined to such a
lattice
is larger than the $2^n$-dimensional computational
Hilbert space $\CC$ which their positions encode.
The encoding is accomplished with a subset $\H_\CC\subset\H$, such
that $|\H_\CC|=2^n$, along with an isomorphism $\enc:\H_\CC\to\CC$.
Generally we can use \CC\ and $\H_\CC$ interchangeably without
ambiguity.
In order to generate
entangled states in \CC\ we make use of a Hamiltonian that couples
$\H_\CC$ to states outside of the computational space, i.e.\ those
in $\H_\perp\equiv\H\setminus\H_\CC$.
Under our scheme, a physical state $\ket{\Psi}\in\H_\CC$
provides a valid encoding of a computational state on $n$ qubits,
$\ket{\Psi}_\CC=\enc(\ket{\Psi})$, while
any state $\ket{\alpha}\in\H$ that has non-zero support in $\H_\perp$
has no such encoding.
Since the computationally entangling Hamiltonian couples $\H_\CC$ to
$\H_\perp$, the evolution of the system restricted to
\CC\ appears to be non-unitary
in that the magnitude of the projection of the physical state of
the system onto the subspace $\H_\CC$ can be different from unity.
However, we show that an arbitrary initial state with support entirely in
$\H_\CC$ will,
at well-defined times in its evolution under the Hamiltonians we
specify, map onto a state that also has no support in $\H_\perp$.
That is, all probability will return to \CC\ and the physical system
will once again  encode a valid computational state.

The ability to construct an entangling Hamiltonian under this encoding
relies on the indistinguishability of the bosons involved
as the encoding we use requires the operations of interchanging two
particles on the lattice and of swapping them twice, returning them to
their initial configuration, to be identical.
This is crucial for preserving the mapping from the physical system
to the computational space that requires one bosonic walker to be
localized to two vertices of the primary graph.

We assume that unless an explicit operation
is being performed, the lattice is deep enough to prevent all tunneling.
In this default configuration the system
Hamiltonian is
\beq
	\Ham_0 = \frac{\onsite}{2}\sum_{i=1}^n\sum_{x=0}^1 \numop_{i,x}
					\left(\numop_{i,x}-1\right).
\eeq
This results in a primary graph that is completely disconnected
--- $2n$ vertices with no edges, $E=\varnothing$.
Each two-site column of the lattice represents a qubit, with the upper
and lower
sites corresponding to the $\cket{0}$ and $\cket{1}$
computational basis states of that qubit, respectively.
An $n$-particle state $\ket{\Psi}$ represents a valid computational
state if and only if
\beq\label{eq:oneParticleCond}
	\sum_{x=0}^1 \left|
		\bra{\Psi}
		\cdop{i,x} \cop{i,x}
		\ket{\Psi}
	\right|^2
	=
	1
\eeq
for each $i$ from
$1$ to $n$, i.e.\ if there is one particle in each column.
Physical states that satisfy this criterion are mapped onto computational ones in a canonical way,
\begin{subequations}\label{eq:oneQubitBasis}
\begin{align}
	\cket{0}_i &\leftrightarrow \cdop{i,0}\ket{00}_i
			= \ket{10}_i, \\
		\text{and}\ 
	\cket{1}_i &\leftrightarrow \cdop{i,1}\ket{00}_i
			= \ket{01}_i.
\end{align}
\end{subequations}
We use underlines to indicate encoded computational states, in contrast
with physical Fock-number states.
The system is initialized in the state
\beq
	\ket{\Psi_0}
	=
	\bigotimes_{i=1}^n c_{i,0}^\dg \vac
	\leftrightarrow
	\cket{0}^{\otimes n}.
\eeq
This is a zero-energy eigenstate of $\Ham_0$,
and as such is stationary while the lattice is maintained
and the primary graph has no edges.

The dimension of \H\ is $\binom{2n}{n}$,
in general much larger than $|\CC|=2^n$.
This may seem inefficient or even detrimental at first sight, but
it is this access to a larger Hilbert space during evolution
that allows for the implementation of a two-qubit controlled gate
through only on-site interactions with no internal degrees of
freedom.  We impose Hamiltonians that
couple $\H_\CC$ to $\H_\perp$ in such a way that at
well-defined times the resulting unitary operators block diagonalize
as $\unitary=\unitary_\CC\oplus \unitary_\perp$.
That is, all population initially in
a valid computational state returns to the computational subspace
at the end of the evolution, despite having been transferred through
the larger space at intermediate times.

Before discussing our two-qubit gates, in particular an entangling
\cphase\ gate and a nearest-neighbor \swap,
we first provide a simple method
for implementing arbitrary single-qubit
operations.

\subsection{Single-qubit operations}

Consider the $i$th encoded qubit, with computational basis states
$\cket{0}_i$ and $\cket{1}_i$ encoded according to
\eqref{eq:oneQubitBasis} in a physical state
$\ket{\Psi}$ satisfying the single-particle condition
\eqref{eq:oneParticleCond}.
The single-qubit operations we construct below trivially preserve this
condition, as they do not couple $\ket{\Psi}$ to states outside of the
computational space.

We first show how to implement an arbitrary $X$ rotation on qubit $i$.
By lowering the height of the lattice barrier between the 
two sites we can set the system Hamiltonian to
\beq\label{eq:HXphysical}
	\Ham_{X,i} = -\hop_{X,i}\left(\cdop{i,0}\cop{i,1} + \Hc\right)
					+ \Ham_0,
\eeq
where $\hop_{X,i}$ depends on the height of the lowered barrier between
the sites.
The primary graph in this case is composed of $2(n-1)$ disconnected
vertices and one copy of $K_2$, the connected two-vertex graph, on
the vertices encoding qubit $i$.
The action of $\Ham_{X,i}$ on the basis states \eqref{eq:oneQubitBasis}
is simply
\begin{subequations}
\begin{align}
	\Ham_{X,i}:\,
		&\ket{10}_j
			\mapsto
			-\delta_{ij}\hop_{X,i}\ket{01}_j, \\
  		&\ket{01}_j
			\mapsto
			-\delta_{ij}\hop_{X,i}\ket{10}_j,
\end{align}
\end{subequations}
where $\delta_{ij}$ is the Kronecker delta.
In the computational space this acts as an $X$ operator on qubit $i$,
\begin{align}
	\Ham_{X,i}:\,
		&\cket{0}_i \mapsto -\hop_{X,i} \cket{1}_i, \\
		&\cket{1}_i \mapsto -\hop_{X,i} \cket{0}_i,
\end{align}
and does nothing to the other qubits encoded in $\ket{\Psi}$.
The unitary operator
generated by evolution under this Hamiltonian for a time $t$ is then
$\unitary_{X,i}(t)=\rot_{X,i}(-2\hop_{X,i}t)$,
an $X$ rotation of the $i$th qubit.

Next we construct a $Z$ rotation on qubit $i$.
To do so, we maintain the initial height of the barrier between the
two sites while applying a local potential
to the $\cket{1}_i$ site.  The physical Hamiltonian becomes
\beq
	\Ham_{Z,i}
	=
	-\applied_{Z,i} \numop_{i,1} + \Ham_0
\eeq
and corresponds to a primary graph of $2n$ disconnected vertices, with
a self-loop attached to the second of the two vertices that encode
qubit $i$.
The Hamiltonian acts only on the $\cket{1}_i$ computational basis state,
as
\beq
	\Ham_{Z,i}:\,
	\cket{1}_j\mapsto -\delta_{ij}\applied_{Z,i}\cket{1}_j,
\eeq
so the resulting unitary is
$\unitary_{Z,i}(t)=e^{i\applied_{Z,i}t/2}\rot_{Z,i}(\applied_{Z,i}t)$
--- a $Z$ rotation of qubit $i$, up to an unimportant overall phase.

\begin{figure}
	\begin{tikzpicture}
		\node at (0.5,0) {$\cdots$};
		\begin{scope}[line width=1pt]
      \draw (2,0.5) -- (2,-0.5);
      \draw (5,-0.5) .. controls (5.5,0) and (5.5,-1)
        .. (5,-0.5);
    \end{scope}
		\foreach \i in {1,2,3,4,5,6} {
  		  \node[vertex] at (\i,0.5) {};
   		\node[vertex] at (\i,-0.5) {};
		}
		\node at (3.5,0) {$\cdots$};
		\node at (6.5,0) {$\cdots$};
		\node at (2,-1) {$i$};
		\node at (5,-1) {$j$};
	\end{tikzpicture}
	\caption{\label{fig:singleQubitGraphs}%
		An $n$-walker graph on $2n$ vertices,
		on which appropriately initialized
		bosonic walkers will undergo $X$ and $Z$ rotations on the
		encoded qubits $i$ and $j$, respectively.
	}
\end{figure}
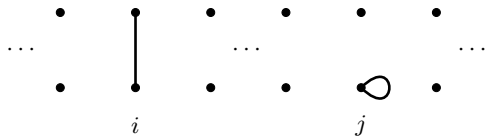

Given an angle $\theta\in[0,2\pi)$ we can therefore perform
$\rot_X(\theta)$ on qubit $i$ by evolving under $\Ham_{X,i}$ for a time
\begin{subequations}
\beq
	t_{X,i}(\theta)=\frac{4\pi-\theta}{2\hop_{X,i}},
\eeq
or $e^{i\theta/2}\rot_Z(\theta)$
by evolving under $\Ham_{Z,i}$ for a time
\beq
	t_{Z,i}(\theta)=\frac{\theta}{\applied_{Z,i}}.
\eeq
\end{subequations}
Given sufficient freedom in the ability to set the tunneling rates and
on-site potentials, it is possible to enact either of these
single-qubit gates on each qubit simultaneously, with different
values of $\theta$ on each one, and have them finish at the same
time. That is, given a set of angles $\theta_i$ and a choice of
gates $\hat{O}_i\in\{\rot_X,\rot_Z,\hat{I}\}$,
the values of $\hop_{X,i}$ and
$\applied_{Z,i}$ can be chosen such that in a fixed time $t$,
the operations
$\hat{O}_i(\theta_i)$ are simultaneously applied across all qubits
$i\in\{1,\ldots,n\}$.
Fig.~\ref{fig:singleQubitGraphs}
shows an example multi-walker primary graph that
applies an $X$ rotation to qubit $i$ and a $Z$ rotation to qubit $j$.
The combination of these operations allows for the execution of
arbitrary single-qubit unitaries in three steps by decomposing
the corresponding rotations of the Bloch sphere using Euler
angles.

Finally, we note that while these single-qubit operators are
sufficient to implement a Hadamard operation on qubit $i$
in three steps, as
\beq
	\Hada_i
	=
	\rot_{X,i}(\pi/2)\rot_{Z,i}(\pi/2)\rot_{X,i}(\pi/2),
\eeq
it is also possible to obtain a Hadamard in a single step with
the Hamiltonian
\beq
	\Ham_{H,i}
	=
	-\applied_{H,i}\numop_{i,0}-\hop_{H,i}
	\left(
		\cdop{i,0}\cop{i,1}+\Hc
	\right)
	+\Ham_0.
\eeq
This simple approach of effectively turning on the Hamiltonians
for $X$ and $Z$ rotations simultaneously results in the application
of a Hadamard gate on qubit $i$, up to an overall phase, at time
\beq
	t_{H,i}=\frac{\pi}{2\sqrt{2}\hop_{H,i}}
\eeq
if the applied-potential--to--hopping ratio is tuned to be
$\applied_{H,i}/\hop_{H,i}=2$.
Not only does this one-step process require fewer dynamical
controls, but for equal hopping and interactions terms
(i.e.\ taking $J_{H,i}=J_{X,i}$ and $V_{Z,i}=V_{H,i}$)
results in a run-time that is an order of magnitude shorter.
This example is unlikely to be the only such shortcut to
additional gates available by judicious choices of further
Hamiltonians.

\subsection{Generating entanglement}
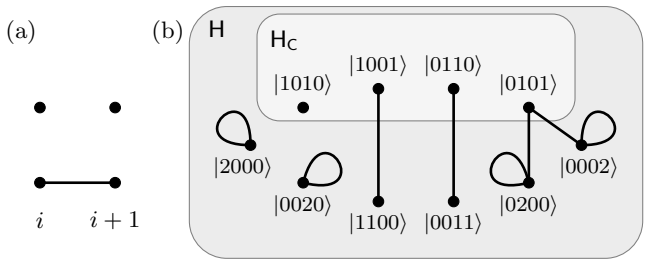
\begin{figure}
	\begin{tikzpicture}[
  			>=stealth,
  			vertex/.style={
			circle,draw,fill=black,
			minimum size=4pt,inner sep=0pt
		}
  		]
		\node at (-5.25,1.5) {(a)};
		\draw[line width=1pt] (-5,-0.5) -- (-4,-0.5);
		\node[vertex] at (-5,0.5) {};
		\node[vertex] at (-5,-0.5) {};
		\node[vertex] at (-4,0.5) {};
		\node[vertex] at (-4,-0.5) {};
		\node at (-5,-1) {$i$};
		\node at (-4,-1) {$i+1$};
		\node at (-3.3,1.5) {(b)};
		\node at (-2.2,0) [vertex] {};
		\node at (2.2,0) [vertex] {};
		\node at (-1.5,-0.5) [vertex] {};
		\node at (-0.5,-0.75) [vertex] {};
		\node at (0.5,-0.75) [vertex] {};
		\node at (1.5,-0.5) [vertex] {};
		\node (v1100) at (-1.5,0.5) [vertex] {};
		\node (v1001) at (-0.5,0.75) [vertex] {};
		\node (v0110) at (0.5,0.75) [vertex] {};
		\node (v0011) at (1.5,0.5) [vertex] {};
		\node (k1100) at (-1.5,0.8)
			{\footnotesize \ket{1010}};
		\node (k1001) at (-0.5,1.05)
			{\footnotesize \ket{1001}};
		\node (k0110) at (0.5,1.05)
			{\footnotesize \ket{0110}};
		\node (k0011) at (1.5,0.8)
			{\footnotesize \ket{0101}};
		\node (compSpace) at (-1.75,1.4) {$\H_\CC$};
		\node (k2000) at (-2.3,-0.3)
			{\footnotesize \ket{2000}};
		\node at (-1.5,-0.8)
			{\footnotesize \ket{0020}};
		\node at (-0.5,-1.05)
			{\footnotesize \ket{1100}};
		\node (k0101) at (0.5,-1.05)
			{\footnotesize \ket{0011}};
		\node at (1.5,-0.8)
			{\footnotesize \ket{0200}};
		\node (k0002) at (2.3,-0.3)
			{\footnotesize \ket{0002}};
		\begin{scope}[line width=1pt]
			\draw (-0.5,0.75) -- (-0.5,-0.75);
			\draw (0.5,0.75) -- (0.5,-0.75);
			\draw (2.2,0) -- (1.5,0.5) -- (1.5,-0.5);
			\draw (-2.2,0) .. controls (-2.2,1) and (-3.2,0) .. (-2.2,0);
			\draw (-1.5,-0.5) .. controls (-1.25,0.5) and (-0.5,-0.75) .. (-1.5,-0.5);
			\draw (1.5,-0.5) .. controls (1.25,0.5) and (0.5,-0.75) .. (1.5,-0.5);
			\draw (2.2,0) .. controls (2.2,1) and (3.2,0) .. (2.2,0);
		\end{scope}
		\node at (-2.65,1.55) {$\H$};
		\begin{pgfonlayer}{background}
			\node[
				inner sep=2mm,
				draw=gray, fill=gray!15, rounded corners=5mm,
				fit = (compSpace.north) (k2000) (k0101) (k0002)
			] {};
			\node[
	    		inner sep=1mm,
	    		draw=gray, fill=gray!8, rounded corners=3mm,
	    		fit = (k1100) (k0011) (v0011)
							(compSpace.north)
			] {};
		\end{pgfonlayer}
	\end{tikzpicture}
	\caption{\label{fig:compSubspace}%
		(a) The primary two-walker subgraph on which encoded qubits $i$
		and $i+1$ undergo a \cphase\ operation has four vertices
		and a single edge.
		(b) The corresponding secondary graph, describing the
  		couplings among Fock states under the Hamiltonian
		$H_\cp$, in columns $i$ and $i+1$ of the physical graph,
		has 10 vertices to represent the 10-dimensional Fock space
		$\H$ of two bosons on four sites. Four of these are
		selected to encode the computational space $\H_\CC$.
		Edges between vertices correspond to allowed
		transitions, and self loops to on-site interactions.
	}
\end{figure}

We now show how to generate entanglement between adjacent qubits,
in the form of a controlled phase gate, $\cphase(\phi)$.  We must expand our
discussion to a $2\times2$ contiguous sub-block of the entire lattice,
which we take to be the vertices $v_{i,x}$ and
$v_{i+1,x}$, for $x\in\{0,1\}$, and we again assume that the 
initial $n$-qubit
state $\ket{\Psi}$ satisfies condition \eqref{eq:oneParticleCond},
containing one bosonic walker per two-site column.
Restricting our attention to two walkers on four sites,
the physical space in question is 10 dimensional.
Four basis states correspond to the computational basis,
and we will couple these to an additional four physical
states.  The remaining two physical basis states can be ignored so long
as the initial state is a computational one.
For convenience we drop subscripts and write the Fock states on
the four vertices in question as a single ket when there is no
ambiguity from doing so, taking
for example $\ket{0110}=\ket{01}_i\ket{10}_{i+1}$.
The two-qubit computational space is encoded in the Fock number states
\begin{subequations}
\begin{align}
	\cdop{i,0} \cdop{i+1,0}\ket{0000}
		&= \ket{1010} \leftrightarrow \cket{00},\\
	\cdop{i,0} \cdop{i+1,1}\ket{0000}
		&= \ket{1001} \leftrightarrow \cket{01},\\
	\cdop{i,1} \cdop{i+1,0}\ket{0000}
		&= \ket{0110} \leftrightarrow \cket{10},\\
	\cdop{i,1} \cdop{i+1,1}\ket{0000}
		&= \ket{0011} \leftrightarrow \cket{11}.
\end{align}
\end{subequations}
The six physical basis states
$\ket{1100}$, $\ket{0011}$, $\ket{2000}$, $\ket{0200}$,
$\ket{0020}$, and $\ket{0002}$ have no computational interpretation.
We can think about the two-qubit situation in two complementary
ways. The first is as we have described, a four-vertex primary
graph on which
there are two interacting quantum walkers.
The second is to treat each of the physical basis states as a vertex
in a new single-walker secondary graph on 10 vertices,
with edges prescribed
by the transitions allowed under the Hamiltonian in question.
For example,
the physical and computational bases under the Hamiltonian
we will introduce for generating entanglement are represented
in Fig.~\ref{fig:compSubspace}, with the primary graph in (a)
and the corresponding secondary graph in (b).

To implement a \cphase\ gate, we decrease
the barrier height between the sites corresponding
to $\cket{1}_i$ and $\cket{1}_{i+1}$ so that
the system Hamiltonian is
\beq
	\Ham_{\cp,i} = 
		-\hop_\cp\left(\cdop{i,1} \cop{i+1,1} + \Hc\right)
		+ \Ham_0.
\eeq
Let $\unitary_{\cp}(t)\equiv\exp(-\imath \Ham_{\cp} t)$
be the time-evolution
operator generated by $\Ham_{\cp}$, where we have dropped the `$i$'
subscripts when no ambiguity arises from doing so.  Then for any
time $t$, the $\cket{00}$ computational state evolves
as $\unitary_{\cp}(t)\ket{1100}=\ket{1100}$.
The state encoding $\cket{01}$ couples outside of the
computational basis, to $\ket{1100}$; at time $t$,
\beq
	\unitary_{\cp}(t)\ket{1001}
	=
	\cos(\hop_\cp t)\ket{1001}+\imath\sin(\hop_\cp t)\ket{1100}.
\eeq
In order to guarantee that $\unitary_\cp$ maps computational states
to computational states, we must therefore require it to act
for a time $t_{\cp,k}\equiv k\pi/\hop_\cp$, $0<k\in\Z$.
This also satisfies the requirement that $\cket{10}$ be
mapped to the computational basis.  Specifically, for
$\ket{\psi}\in\{\ket{1001},\ket{0110}\}$ we have
$\unitary_\cp(t_{\cp,k})\ket{\psi}=(-1)^k\ket{\psi}$.

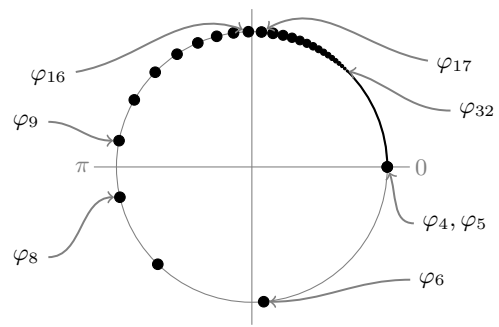
\begin{figure}
	\begin{tikzpicture}[scale=0.6]
		\draw[gray] (-3.5,0) -- (3.5,0);
		\draw[gray] (0,3.5) -- (0,-3.5);
		\draw[gray] (0,0) ellipse (3 cm and 3 cm);
		\foreach \k in {4,...,17} {
			\node[circle,draw=black,fill=black,
						minimum size=4pt,inner sep=0pt]
				({phase\k})
				at ({-pi*(\k+sqrt(\k*\k-16)) r}:3) {};
		}
		\foreach \k in {18,...,32} {
			\node[circle,draw=black,fill=black,
						minimum size=(4-3.5*(\k-18)/15),inner sep=0pt]
				({phase\k})
				at ({-pi*(\k+sqrt(\k*\k-16)) r}:3) {};
		}
		\draw[thick] (3,0) arc (0:45:3);
		\node (phi32) at (5,1.25) {$\varphi_{32}$};
		\node (phi17) at (4.5,2.25) {$\varphi_{17}$};
		\node (phi16) at (-4.5,2) {$\varphi_{16}$};
		\node (phi9) at (-5,1) {$\varphi_9$};
		\node (phi8) at (-5,-2) {$\varphi_8$};
		\node (phi6) at (4,-2.5) {$\varphi_6$};
		\node (phi4) at (4.5,-1.25) {$\varphi_4,\varphi_5$};
		\tikzset{labelArrow/.style =
			{->,line width=0.75pt,gray}
		}
		\draw[labelArrow] (phi32) to[out=180,in=345] (phase32);
		\draw[labelArrow] (phi17) to[out=170,in=30] (phase17);
		\draw[labelArrow] (phi16) to[out=10,in=160] (phase16);
		\draw[labelArrow] (phi9) to[out=0,in=180] (phase9);
		\draw[labelArrow] (phi8) to[out=0,in=180] (phase8);
		\draw[labelArrow] (phi6) to[out=180,in=0] (phase6);
		\draw[labelArrow] (phi4) to[out=180,in=290] (phase4);
		\node[gray] at (3.75,0) {$0$};
		\node[gray] at (-3.75,0) {$\pi$};
	\end{tikzpicture}
	\caption{\label{fig:phases}%
		Distribution of available phases $\varphi_m$ satisfying
		\eqref{eq:phim}. Except for the two trivial values,
		$\varphi_4$ and $\varphi_5$, no phase is a rational multiple
		of $\pi$.
		For large $m$ the phase goes as $-8\pi/m$, modulo $2\pi$,
		returning to 0 as $m\to\infty$, which corresponds to the limit
		of zero hopping.
	}
\end{figure}

The fourth computational state on two qubits, $\cket{11}$,
evolves within a three-dimensional subspace spanned by
$\ket{0101}$, $\ket{0200}$, and $\ket{0002}$, as illustrated by
the largest connected component of the secondary graph in
Fig.~\ref{fig:compSubspace}(b). 
Determining the action of $\unitary_\cp$ on $\ket{0101}$ shows
that for a walker initially in this state to return entirely to the 
computational basis at time $t_{\cp,k}$ requires
\beq
	\exp\left[2\pi \imath\sqrt{16+\frac{\onsite^2}{\hop_\cp^2}}\right]
	=
	1.
\eeq
This is accomplished if the ratio of the on-site repulsion to
the hopping rate is tuned to be
$\onsite/\hop_\cp=\sqrt{m^2-16}$,
for any integer $m\ge4$.
The final requirement to guarantee that $U_\cp$ maps the computational
space onto itself at time $t_{\cp,k}$ is
that the product $mk$ be even.
We can guarantee this by setting $k=2$, defining
$t_\cp=2\pi/J_\cp$.
In this case the action of $\unitary_\cp$
decomposes into its action on the computational basis states,
and its action on the rest of Fock space. In the canonical ordering of
the computational basis states, we have
\begin{subequations}\label{eq:CPHASE}
\begin{align}
	\unitary_\cp(t_\cp)
	&=
	\diag\left(1,1,1,e^{-\imath\pi\left(m+\sqrt{m^2-16}\right)}\right)
		\oplus \unitary_\cp^\perp \\
	&=
	\cphase(\varphi_m)
		\oplus \unitary_\cp^\perp,
\end{align}
\end{subequations}
where $\unitary_\cp^\perp$ acts only on the subspace orthogonal to
the computational one, and
\beq\label{eq:phim}
	\varphi_m\equiv-\pi\left(m+\sqrt{m^2-16}\right).
\eeq
This is a non-trivial entangling phase for any  $m>5$, and
the available values are depicted graphically
on the unit circle in Fig.~\ref{fig:phases}.

If we instead choose an odd value for $k$, including $k=1$ which
results in a shorter run-time for the gate, the even values of
$m\ge4$ result in the same set of phases but indexed by $m/2$
rather than $m$. The resulting
gate in this case is $(Z\otimes Z)\cphase(\varphi_{m/2})$.

\subsection{\uppercase{Swap} gate}
In order to provide a universal gate set with only a nearest-neighbor
entangling gate, a two-qubit \swap\ gate is also required.
A variation of the $X$ rotation performed by $\Ham_{X,i}$
allows for a straightforward implementation of this by the
simultaneous lowering of the potential barrier between sites
$\cket{x}_i$ and $\cket{x}_{i+1}$, for each $x\in\{0,1\}$.
The system Hamiltonian in this case is
\beq
	\Ham_{S,i}
	=
	-\hop_{S}\left(
		c_{i,0}^\dg c_{i+1,0}^{\vphantom{\dg}}
		+c_{i,1}^\dg c_{i+1,1}^{\vphantom{\dg}}
		+\Hc
	\right)
	+\Ham_0.
\eeq
This acts non-trivially on the four basis states satisfying
\eqref{eq:oneParticleCond}, coupling the computational space
to all six of the remaining physical basis states.
As with the analysis of the \cphase\ gate, we can place restrictions
on the available parameters such that the action of the operation
restricted to the computational space is unitary at the end of the
evolution. The action of $\Ham_{S}$ on
$\ket{1001}$ and $\ket{0110}$ requires that we set the \swap\
time to be
\begin{subequations}\label{eq:swapConds}
\beq
t_{S,k}=\frac{(2k+1)\pi}{2\hop_S},
\quad
0\le k\in\Z.
\eeq
Under this restriction, the action of
$\Ham_S$ on $\ket{1010}$ and $\ket{0101}$ further requires that
\beq
	\frac{\onsite}{\hop_S}
	=
	4\sqrt{\frac{l^2}{(2k+1)^2}-1},
	\quad
	2k+1<l\in\Z.
\eeq
\end{subequations}
When these conditions are satisfied, the action of
$\unitary_S(t)\equiv\exp(-\imath\Ham_S t)$ at time $t=t_{S,k}$
block diagonalizes such that its effect on the computational
space is that of
\beq\label{eq:swapUnitary}
	\left.\unitary_S(t_{S,k})\right|_{\H_\CC}
	=
	\begin{pmatrix}
		e^{-\imath\alpha\pi} & 0 & 0 & 0 \\
		0 & 0 & -1 & 0 \\
		0 & -1 & 0 & 0 \\
		0 & 0 & 0 & e^{-\imath\alpha\pi}
	\end{pmatrix}
\eeq
with
\beq
	\alpha=l+\sqrt{l^2-4k(k+1)-1}.
\eeq
In general this provides us with a second entangling gate,
unless $\alpha$ is an integer.
In that case, if $\alpha$ is even then \eqref{eq:swapUnitary}
is equivalent to $(Z\otimes Z)\swap$, and if it is odd then
we obtain the gate $-\swap$.

The problem of finding values
for $k$ and $l$ 
that satisfy the conditions
\eqref{eq:swapConds} while resulting in an integral value of
$\alpha$
reduces to finding Pythagorean triples, of which
there are infinitely many.
One such combination, which results in the minimal
time for the operation, is $k=1$ and $l=5$. This
pair leads to the parameters
$t_S = 3\pi/(2\hop_S)$ and $\onsite/\hop_S = 16/3$,
and implements a two-qubit nearest-neighbor \swap\ gate, up to an
overall phase, as required for a universal gate set.

\subsection{Connection to discontinuous walks}
Implementing a sequence of computational gates requires that we
toggle a set of potentials on and off in a prescribed order,
affecting the evolution of a set of quantum walkers in the process.
This is in the same spirit as the single-walker discontinuous
walk we have previously proposed \cite{Underwood2010-PRA82-42304},
though that work makes use of a number of vertices exponential in the
number of encoded qubits whereas in the current multi-walker
scheme only linear growth is required in the number of vertices.

\begin{figure}
	\begin{tikzpicture}[x=0.333cm,y=0.333cm]
		%
		%
		\node at (-5.5,10) {(a)};
		\latticeTimestep{0}{8}{a}
		\latticeTimestep{7}{8}{c}
		\latticeTimestep{14}{8}{b}
		\node at (-3.5,10) {$\scriptstyle\ket{0}:$};
		\node at (-3.5,8) {$\scriptstyle\ket{1}:$};
		\draw (sitea10) -- (sitea11);
		\draw (siteb10) -- (siteb20);
		\draw (siteb11) -- (siteb21);
		\draw (sitec11) -- (sitec21);
		%
		%
		\node at (0,5.25) {$\rot_X\otimes I$};
		\node at (7,5.25) {$\cphase$};
		\node at (14,5.25) {$\swap$};
		%
		%
		\node at (-5.5,4.5) {(b)};
		\railTimestep{0}{0}{a}
		\railTimestep{7}{0}{c}
		\railTimestep{14}{0}{b}
		\foreach \y in {0,...,3} {
			\draw[gray, line width=0.5pt] (-4,\y) -- (18,\y);
		}
		\node at (-5,3) {$\scriptstyle\ket{00}$};
		\node at (-5,2) {$\scriptstyle\ket{01}$};
		\node at (-5,1) {$\scriptstyle\ket{10}$};
		\node at (-5,0) {$\scriptstyle\ket{11}$};
		\draw (va1) to[out=230,in=130] (va3);
		\draw (va2) to[out=310,in=50] (va4);
		\draw (va5) -- (va6) -- (va7);
		\draw (va8) -- (va9) -- (va10);
		\draw (vb5) -- (vb1) -- (vb10);
		\draw (vb2) -- (vb6) -- (vb3) -- (vb9) -- (vb2);
		\draw (vb7) -- (vb4) -- (vb8);
		\draw (vc2) -- (vc6);
		\draw (vc3) -- (vc9);
		\draw (vc7) -- (vc4) -- (vc8);
	\end{tikzpicture}
	\caption{\label{fig:discontinuous}%
	  (a) Hopping parameters turned on to enact gates with two
	  qubits present.
		(b) The corresponding graphs encoding couplings among the
		ten states of $\H$ are presented
		in gray for each operation, with the four computational states of
		$\H_\CC$ highlighted in light gray.
	}
\end{figure}
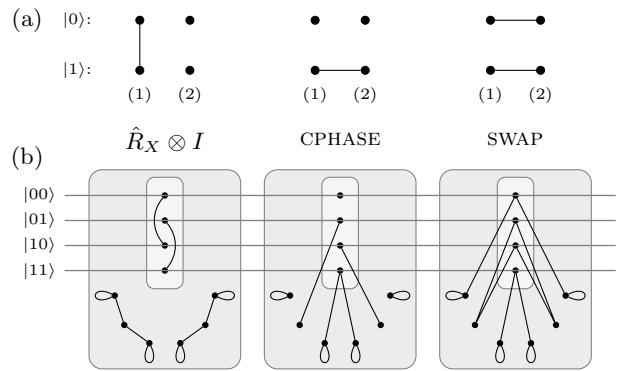

\begin{figure}
	\begin{tikzpicture}[x=0.333cm,y=0.333cm]
		\node at (-15, 6) {(a)};
		\node at (-14, 2.5) {$\scriptstyle\ket{0}:$};
		\node at (-14, 0.5) {$\scriptstyle\ket{1}:$};
		\foreach \x / \lblx in {1/3,2/2,3/1} {
		  \node at (-6-2*\x, -1) {$\scriptstyle(\lblx)$};
		  \foreach \y / \lbly in {0/1,1/0} {
		    \node[vertex] (site\lblx\lbly) at (-6-2*\x, 0.5+2*\y) {}; 
		  }
		}
		\draw (site10) -- (site11);
		\draw (site20) -- (site30);
		\draw (site21) -- (site31);
		\node at (-10,4.5) {$\rot_X\otimes\swap$};
		\node at (-4.75, 6) {(b)};
		\swapGraph{0}{0}{a}
		\swapGraph{5}{3.}{b}
		\tikzstyle{joinedge} = [draw];
		\draw[joinedge] (va-1100) -- (vb-1100);
		\draw[joinedge] (va-1001) -- (vb-1001);
		\draw[joinedge] (va-0110) -- (vb-0110);
		\draw[joinedge] (va-0011) -- (vb-0011);
		\draw[joinedge] (va-0101) -- (vb-0101);
		\draw[joinedge] (va-1010) -- (vb-1010);
		\draw[joinedge] (va-2000) -- (vb-2000);
		\draw[joinedge] (va-0200) -- (vb-0200);
		\draw[joinedge] (va-0020) -- (vb-0020);
		\draw[joinedge] (va-0002) -- (vb-0002);	
	  	\begin{pgfonlayer}{background}
			\node[
				inner sep=1mm,
				draw=gray, fill=gray!15, rounded corners=5,
				fit = (bla) (trb)
			] {};
		\end{pgfonlayer}
	\end{tikzpicture}
	\caption{\label{fig:threeQubits}%
	  When three qubits are encoded, (a) three edges among the
	  six vertices of the primary graph
	  implement an $\rot_X\otimes\swap$ gate.
	  (b) The corresponding secondary
	  graph is much larger, and is given by
	  the cartesian product of the one- and two-qubit graphs
	  for $\rot_X$ and \swap.
	  Note that those
	  vertices which are not connected to the computational states
	  have been omitted in an attempt at clarity.
	}
\end{figure}
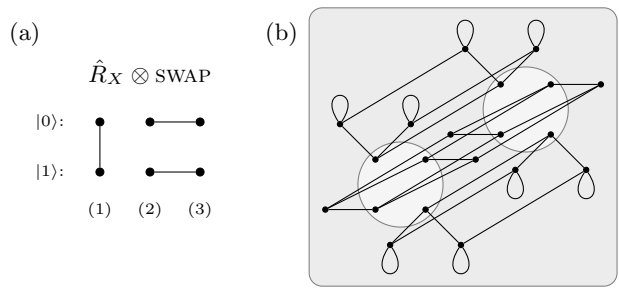

As in the prior models of universal computation by quantum walk
\cite{Childs2009-PRL102-180501,Lovett2010-PRA81-42330},
the single-walker discontinuous model employs a set of
`rails' for encoding the quantum information. These rails are
linear graphs that allow propagation of information from left
to right, interspersed with so-called `widgets' that enact
transformations on the state of the walker as it passes them.
There is one rail for each computational basis state, which leads
to the exponential growth in vertices.
In Fig.~\ref{fig:discontinuous} we show how a time sequence of
two-walker primary graphs on four vertices corresponds to a single
discontinuous walker on a larger secondary graph, and draw a connection
to the rail model.
Since $2n=2^n$ for $n=2$ the number of rails
is equal to the number of vertices in this case, though in
Fig.~\ref{fig:threeQubits} we show what a single step of a
discontinuous walk on three qubits looks like, on six
vertices in the primary multi-walker case and on eight rails in a
56-dimensional space for a single walker on the resulting secondary
graph.

\section{Measurements and error correction}\label{sec:MeasureQECC}
Under a single-walker model for computation,
a measurement of the computational states of $m\le n$ qubits
can be accomplished either in a single step with a set of
$\Omega(2^m)$ physical measurement operators, each spatially
delocalized over $\Omega(2^{n-m})$ vertices, or as a sequence
of $m$ single-qubit measurements, each requiring two
measurement operators delocalized over $\Omega(2^{n-1})$ vertices.
Assume the best-case scenario of there being a one-to-one mapping
between the computational basis
states and a (sub)set of $2^n$ vertices.
For $x\in\{0,\ldots,2^{m}-1\}$, let $x=x_1\cdots x_m$ be
the $m$-bit binary expansion of $x$. To make a measurement
of qubits $\{i_k\}_{k=1}^m$, let
$\mathcal{V}_x$ be the set of integers between $0$ and
$2^n-1$ whose $n$-bit binary expansions have bit value $x_k$
at bit position $i_k$. Then the measurement operators required
are
\beq
	\left\{
		P_x = \sum_{v\in\mathcal{V}_x}\ketbra{v}{v}
	\right\}_{x=0}^{2^m-1}.
\eeq
Clearly there are $2^m$ values of $x$, and for each $x$ there
are $2^{n-m}$ unspecified bit values so
$|\mathcal{V}_x|=2^{n-m}$.
To implement $m'$ single-qubit measurements, simply set $m=1$
and repeat $m'$ times.
This exponential growth in the spatial extent of the required
measurement operators
is why previous quantum-walk models have been proposed
primarily in terms of computational capability, and not in terms
of possible physical implementations.
It also makes the prospect of implementing quantum error
correcting codes impractical at best.

In contrast, the \BH-based multi-walker model
presented here allows
for the measurement of $m$ qubits with localized measurement
operators.
Define
\beq
	P_{i,x}^{(b)}
	=
	\big(\cdop{i,x})^b\ketbra{0_{i,x}}{0_{i,x}}\big(\cop{i,x}\big)^b
	=
	\ketbra{b_{i,x}}{b_{i,x}},
\eeq
the projector onto the state with exactly $b$ walkers on vertex
$v_{i,x}$. The qubit encoded on vertices $v_{i,0}$ and $v_{i,1}$
can then be measured with the operators
\begin{subequations}\label{eq:multiMeasOps}
\begin{align}
	M_{i,0} &= P_{i,0}^{(1)}\otimes P_{i,1}^{(0)}\otimes I_{\bar{i}}, \\
	M_{i,1} &= P_{i,0}^{(0)}\otimes P_{i,1}^{(1)}\otimes I_{\bar{i}},
\end{align}
\end{subequations}
where $I_{\bar{i}}$ is the identity operator on every vertex not
labeled by $i$.
These two operators form a measurement basis for the system
since the allowed states when a measurement is to be performed have
at most one walker on vertex $v_{i,x}$, so the identity operator on
that vertex can be resolved as
\beq
	I_{i,x}
	=
	\ketbra{0_{i,x}}{0_{i,x}}+\ketbra{1_{i,x}}{1_{i,x}}
	=
	P_{i,x}^{(0)}+P_{i,x}^{(1)}.
\eeq
Furthermore the probability to find either zero or two (or more)
walkers on the pair of vertices $\{v_{i,0},v_{i,1}\}$ vanishes, so
all other two-site projectors
\beq
	P_{i,0}^{(a)}\otimes P_{i,1}^{(b)},
	\quad a+b\neq 1,
\eeq
are not part of the basis for the states encoding qubit $i$.

As in the single-walker case we can perform a single-qubit measurement
with one pair of measurement operators, but in this case each operator
is spatially localized to two vertices regardless of $n$.
Performing an $m$-qubit measurement in a
single step, as opposed to by making $m$ single-qubit measurements
in succession, requires a set of $2^m$ measurement operators of the
form \eqref{eq:multiMeasOps}. Again, each one requires projectors
only onto single vertices.

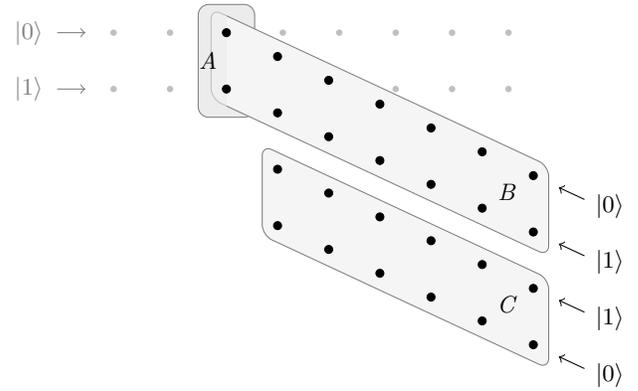
\begin{figure}
	\begin{tikzpicture}[x=0.75cm,y=0.75cm,z={(0.68cm,-0.317cm)}]
		%
		\foreach \x in {-2,...,5} {
		  \node[smlvtx,gray!50] at (\x,0,0) {};
		  \node[smlvtx,gray!50] at (\x,1,0) {};
		}
		\node[gray] at (-3.5,0) {$\ket{1}$};
		\node[gray] at (-3.5,1) {$\ket{0}$};
		\draw[->,gray] (-3,0) -- (-2.5,0);
		\draw[->,gray] (-3,1) -- (-2.5,1);
		%
		\path[draw=gray,fill=gray!15,rounded corners]
    	  (-0.5,-0.5) -- (-0.5,1.5) -- (0.5,1.5)
	    -- (0.5,-0.5) -- cycle;
		\path[draw=gray,fill=gray!8]
			(0,1.3,0) [rounded corners] --  (0,1.3,6.3)  -- (0,-0.3,6.3)
			  [rounded corners=0cm] -- (0,-0.3,0);
		\path[draw=gray!50,fill=gray!13]
			(0,1.3,0) [rounded corners] -- (0,1.3,-0.3) -- (0,-0.3,-0.3)
			  [rounded corners=0cm] -- (0,-0.3,0);
		\path[draw=gray,fill=gray!8,rounded corners]
			(0,-0.7,0.7) -- (0,-0.7,6.3) -- (0,-2.3,6.3)
			  -- (0,-2.3,0.7) -- cycle;
		%
		\foreach \z in {1,...,6} {
			\foreach \y in {-2,...,1} {
				\node[vertex] at (0,\y,\z) {};
			}
		}
		%
		\node[vertex] at (0,0,0) {};
		\node[vertex] at (0,1,0) {};
		%
		\foreach \y in {1,...,-2} {
		  \draw[->] (0,\y,7) -- (0,\y,6.5);
		}
		\node at (0,1.1,7.5) {$\ket{0}$};
		\node at (0,0.1,7.5) {$\ket{1}$};
		\node at (0,-0.9,7.5) {$\ket{1}$};
		\node at (0,-1.9,7.5) {$\ket{0}$};
		%
		\node at (-0.325,0.5) {\emph{A}};
		\node at (0,0.5,5.5) {\emph{B}};
		\node at (0,-1.5,5.5) {\emph{C}};
	\end{tikzpicture}
	\caption{\label{fig:QECC}%
		Addition of the seven-qubit Steane quantum error-correcting
		code to the \BH-based multi-walker scheme for
		universal quantum computation.
		The computation scheme described in
		Sec.~\ref{sec:PositionStates}
		takes place on the horizontal row of light-gray vertices,
		in which the two in region \emph{A} have been singled out
		as the qubit to be encoded in this cartoon.
		The 14 vertices of region \emph{B}
		(including those in \emph{A}) provide seven physical qubits
		with which to encode a single logical qubit.
		Region \emph{C} provides six additional ancillary qubits
		that can be entangled with those of the logical qubit in order
		to perform syndrome measurements as a sequence of single-qubit
		measurements.
		Note that the ordering of the $\ket{0}$ and $\ket{1}$ rows
		has been reversed in \emph{C} with respect to \emph{B};
		this allows a straightforward method of generating entanglement
		between the two regions by way of the \cphase\ gate that
		results from an edge between the $\ket{1}$ vertices of
		neighboring qubits.
	}
\end{figure}

With the ability to perform single-qubit measurements on the
multi-walker system in a straightforward manner
comes the ability to implement those quantum error correcting codes
(QECCs)
that rely on encoding a single logical qubit in multiple physical
ones
\cite{%
Shor1995-PRA52-2493,%
Calderbank1996-PRA54-1098,%
Laflamme1996-PRL77-198%
}.
One such code is the seven-qubit Steane code
\cite{Steane1996-PRSLA1954-2551}, a so-called CSS code
capable of detecting and correcting arbitrary single-qubit errors
and of being implemented fault tolerantly since operations on
encoded logical qubits can be implemented by way of local operations
on the underlying physical qubits.
This and other QECCs can be implemented directly within the
multi-walker framework already discussed in previous sections,
constructing the requisite operators to within a desired error
tolerance from a polynomial number of gates from the universal
set provided.
An alternative,
which we now discuss for the seven-qubit Steane code in particular,
is to use a straightforward extension of the qubit-encoding scheme.
Consider extending the $2\times n$ grid
into the third dimension, to a
$2\times7\times n$ array of vertices or, when it is easier
to perform syndrome measurements as single-qubit
measurements rather than the four-qubit measurement operators
described by the code,
to $4\times7\times n$, as depicted in Fig.~\ref{fig:QECC}.
Each pair of vertices encoding a single qubit in the original
scheme described above, as in Region \emph{A} of Fig.~\ref{fig:QECC},
has a further six pairs extending from it
along the third dimension of the lattice. The resulting total of
seven pairs at a given position $i$ within the lattice,
Region \emph{B}, allows
for the creation of a single logical qubit. Six further
ancillary qubits, Region \emph{C},
can be entangled with the physical qubits of the logical one
in order to implement syndrome measurements on single qubits.
Making the vertices that correspond to the $\ket{1}$ states in
the two regions adjacent allows entanglement to be generated
between them by the \cphase\ gate \eqref{eq:CPHASE} with the
straightforward addition of edges between the logical region
and the ancillary one.

Logical operations on these additional qubits can of course be
performed in exactly the same manner as on the original qubits
described in Sec.~\ref{sec:PositionStates}.
Those gates
are dependent on certain couplings between pairs of vertices,
but are not inherently reliant on a grid- or lattice-like
structure in the placement of those vertices; such a setting
merely provides a useful visualization method, and yields an obvious
tie to possible extant physical systems as we discuss in
Sec.~\ref{sec:Implementation}.
They can therefore all be applied between pairs of qubits in this
new arrangement as well.

Having seen that our multi-walker \BH-based approach to
universal computation provides the advantage of being able to
implement computational error correction, previously unaddressed
within quantum walk-based quantum computation, we also note that
the dynamical nature of the underlying graph introduces a second
potential source of error.
If a gate is timed incorrectly, it is
possible for the system to end up in a physical state that does not
encode a computational one -- besides being flipped and dephased,
a qubit can be lost altogether.
When this possibility is allowed for, the measurement operators
\eqref{eq:multiMeasOps} no longer sum to the identity and we must
introduce a third operator,
\beq
	M_{i,\text{err}}
	=
	I-M_{i,0}-M_{i,1}.
\eeq
This provides a simple method for detecting the loss of a qubit,
but not a mechanism for recovering from it.
The possibility of losing the walker exists in the original
proposal for universal quantum computation by a single quantum
walker \cite{Childs2009-PRL102-180501}, and a similar timing issue
is present in the discontinuous-walk scheme
\cite{Underwood2010-PRA82-42304} as well as any potential
physical scheme for the simulation of a discrete-time quantum walk
with a continuously evolving physical system. These are not fundamental
limitations on quantum walk-based computing, but must be considered
in any serious attempt to engineer such systems.

\section{Physical implementation}\label{sec:Implementation}
A key feature required of a physical system that is to implement this
proposal is the ability to provide the
specified $2\times n$ lattice --- or three-dimensional lattice, if
that method of implementing a QECC is chosen ---
such that the tunneling amplitude between neighboring sites is
close to zero when no gate is being enacted.  This corresponds to
the qubits' having a long coherence time.
In practice this can be accomplished in either the 2D or 3D case
by creating a three-dimensional
lattice with isolated wells, and ignoring any
vertices outside of the primary graph to be
implemented.

The additional requirements
of state preparation, manipulation, and read-out
are common to any quantum computer \cite{DiVincenzo1997},
and we discuss them here in the context of our proposal.
State preparation is accomplished by the
initialization of the $\ket{0}^{\otimes n}$
state, by loading one boson onto the first of each pair of sites
in the primary graph.
To manipulate the encoded qubits, it must be possible
to selectively increase the tunneling amplitude  between
given adjacent sites in order to enact $X$, \cphase, and \swap\ gates,
and to change on-site potentials to enact $Z$ gates.
Finally to read out the result of a computation it must be possible
to measure the positions of the bosonic walkers within the lattice,
as discussed in Sec.~\ref{sec:MeasureQECC}.

The experimental scheme proposed in
Ref.~\cite{Mompart2003-PRL90-147901}
includes many of the features required, though it makes use of adiabatic
processes for gate executions.  A combination of this approach with
the sudden potential-landscape changes discussed in
Ref.~\cite{Calarco2000-PRA61-22304} is more appropriate to the
current scheme, and there have been significant
experimental advances in the intervening years
that offer the promise of a proof-of-principle implementation.

There is an obvious connection between our scheme and optical-lattice
experiments, and several experiments satisfy the above
requirements.
One option is to use a liquid-crystal display (LCD)
as a spatial light modulator \cite{Bergamini2004-JOSAB21-1889}.
Such devices
generate arrays of microtraps holographically based on the pattern
of opacity and transparency present on the easily programmed LCD
screen.
The traps have been used to store single neutral atoms per site,
address individual sites, and measure the locations of trapped atoms
within the lattice \cite{Bergamini2004-JOSAB21-1889}, thus
providing a means to create the primary graph as well as implement
preparation, manipulation, and read-out.

Another possibility is to combine a set of recently demonstrated
experimental capabilities.
Wide lattice spacings on the order of \SI{5}{\micro\meter} have been
achieved \cite{Nelson2007-NP3-556},
providing a long coherence time to the sites and
effectively approximating the infinitely deep lattice of the
primary graph.
A quantum gas microscope, employing a high-numerical-aperture lens,
has been used to image individual sites in a traditional optical 
lattice \cite{Bakr2009-N462-74}.
Repurposing such a system to focus a laser to a similar
resolution would provide a method of manipulating qubits by
addressing single sites in the case
of a $Z$ gate, or modifying the potential between sites in the case
of $X$ and \cphase\ gates.
Most recently, arbitrary configurations of atom positions within
a lattice have been implemented \cite{Weitenberg2011-N471-319},
which in particular would allow for the
straightforward preparation of the initial $\ket{0}^{\otimes n}$ state
as a single straight line, one atom wide.
In each of these experiments, read-out of the positions of the
trapped atoms is also performed.

\section{Conclusions}\label{sec:Conclusions}
The \BH\ model, a well-known and widely
applicable description of bosons confined to a lattice, can be used
to generate a universal set of quantum logic gates.  These gates act
on quantum information encoded in the positional states of spinless
bosons, in contrast to the standard method of using internal degrees
of freedom to store information.  Despite arising from Hamiltonians
that couple outside of the computational space, the gates are
nevertheless unitary when the evolution parameters are tuned
appropriately.
The concept of encoding a computational space as a subspace of a
larger physical Hilbert space is often considered to be detrimental
or artificial when couplings are present between the subspace
and the larger space.  We have shown that such a system can gain power
from its ability to access that larger space during evolution, while
nevertheless remaining restricted to the computational subspace by
the end of its evolution.

Bosons hopping on a lattice under the \BH\ model can also
be interpreted as multiple interacting quantum walkers on a primary
graph that encodes the sites and tunneling amplitudes of the lattice
in its vertices and edges, respectively.
The use of multiple quantum walkers allows for a sequence of graphs
on $O(n)$ vertices to encode a quantum computation on $n$ qubits
by discontinuous walk. Such a setup can also be interpreted as a
single quantum walker on a graph with $\Omega(2^n)$
vertices and connected
to the standard rail model for computation by quantum walk,
showing the power of multiple walkers to eliminate the need for
exponentially growing resources.
Furthermore, localized single-qubit
measurement operators are now possible in a quantum-walk scenario,
and we have presented one example of a scheme to implement quantum
error-correcting codes with multiple quantum
walkers under the \BH\ model.

Finally, we have discussed the possibility of adapting current
experimental methods to implement a proof-of-principle
version of our proposal. Trapped neutral atoms in optical lattices
with various methods for addressing individual sites and manipulating
the potential landscape fulfill the requirements of such an
implementation.

\section*{Acknowledgments}

This work was supported by the
Natural Sciences and Engineering Research Council of Canada (NSERC)
and by Alberta Innovates Technology Futures.

\end{document}